# Mapping nanoscale hotspots with single-molecule emitters assembled into plasmonic nanocavities using DNA origami

Rohit Chikkaraddy[1], V. A. Turek[1], Nuttawut Kongsuwan[2], Felix Benz[1], Cloudy Carnegie[1], Tim van de Goor[1], Bart de Nijs[1], Angela Demetriadou[2], Ortwin Hess[2], Ulrich F. Keyser[1], and Jeremy J. Baumberg[1]

[1] NanoPhotonics Centre, Cavendish Laboratory, Department of Physics, JJ Thompson Avenue, University of Cambridge, Cambridge, CB3 0HE, United Kingdom
[2] Blackett Laboratory, Prince Consort Road, Imperial College London, London SW7 2AZ, UK

ABSTRACT.
**Fabricating nanocavities in which optically-active single quantum emitters are precisely positioned, is crucial for building nanophotonic devices. Here we show that self-assembly based on robust DNA-origami constructs can precisely position single molecules laterally within sub-5nm gaps between plasmonic substrates that support intense optical confinement. By placing single-molecules at the center of a nanocavity, we show modification of the plasmon cavity resonance before and after bleaching the chromophore, and obtain enhancements of ≥ 4×10$^3$ with high quantum yield (≥ 50%). By varying the lateral position of the molecule in the gap, we directly map the spatial profile of the local density of optical states with a resolution of ±1.5 nm. Our approach introduces a straightforward non-invasive way to measure and quantify confined optical modes on the nanoscale.**

Coherent coupling of light and single-molecules at room temperature is one of the fundamental goals of nano-optics that would enable widespread adoption as a building block of nanophotonic devices. Over the past decade many constructs have been designed to enhance the coupling of single-emitters, such as planar interfaces,[1] near-field probes,[2] photonic crystals,[3,4] microcavities[5] and metal nanostructures.[6–8] Improved optical coupling results in Purcell enhancement of the emission rate, enhancing efficiency and both spatial and temporal mode-matching of single photon emission.[9–11] Further improvement eventually leads to the strong-coupling regime where nonlinearities can reach the single photon level, an essential characteristic for many quantum devices. So far however most single-emitter strong-coupling required cooling of the system[12–15] increasing cost and complexity. For room temperature nano-optics, plasmonic nanocavities have gained tremendous interest due to their enhanced field confinements.[7,16–18] Integrating optically-active materials (such as molecules, quantum dots, monolayer semiconductors, or diamond vacancy centers) into these cavities is of great importance to access the desired coherent interaction between optical field and exciton. Either the cavity must be fabricated around randomly located emitters such as quantum dots or nitrogen-vacancy centres,[14,19,20] or emitters are randomly located inside the cavity.[21] In realizing the promising hopes for molecules in plasmonic cavities, the major hurdles are: (i) robust assembly of plasmonic nanocavities with reliable nanogaps ($d<5$ nm), and (ii) precise integration of single-molecules into such cavities with a high degree of spatial control.

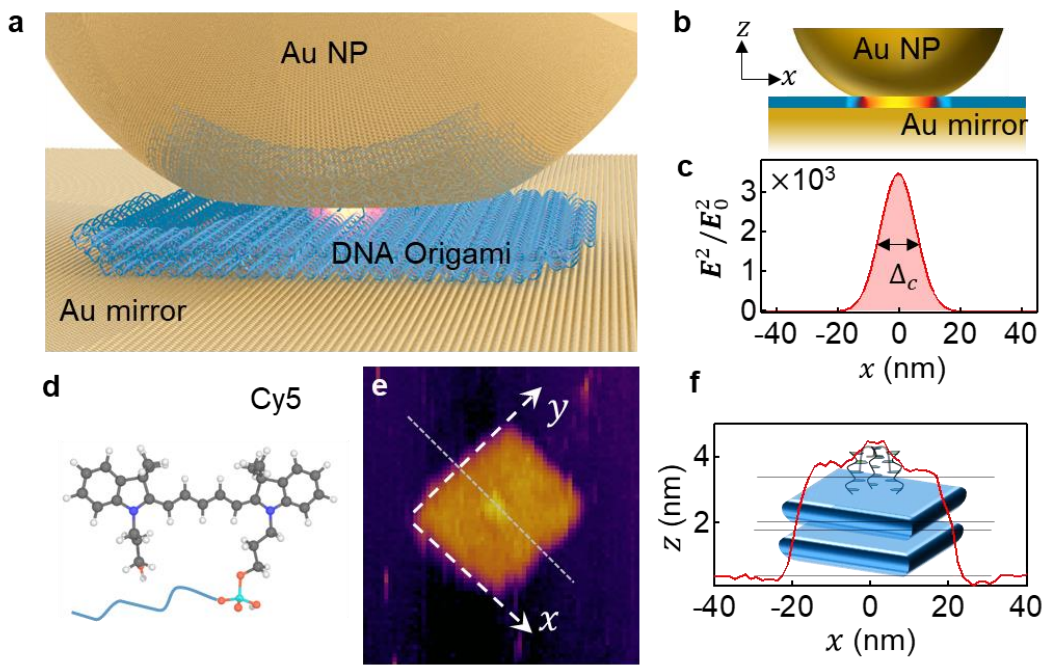

**Figure 1.** Assembled plasmonic nanocavity with single-molecule DNA origami plates. (a) NPoM with facetted nanoparticle and DNAo in the gap. (b) Nanoparticle-on-mirror (NPoM) cavity with strong optical field confinement in the gap (yellow). (c) Electric field enhancement in the gap along the $x$-direction, 80 nm diameter NP with 15 nm bottom facet. (d) Chemical structure of Cy5 molecule positioned in the NPoM gap. (e) Zoomed AFM image of a single DNAo. (f) Line profile along the dotted white line shown in (e). Sketch not to scale.

Here we construct a nanocavity with <5 nm gap between two plasmonic components, and show the freedom to place a single emitter at controlled positions inside it (Fig. 1a). The cavity consists of a gold spherical nanoparticle placed on top of a Au film coated with a robust nanoscale spacer, forming a nanoparticle-on-mirror (NPoM) construct.[22,23] The charge oscillations in the Au nanoparticle couple with image charges within the polarizable Au surface underneath. This enhances the electromagnetic field in the gap by nearly two orders of magnitude and tightly confines the fields to spatial volumes $V_c <$(6 nm)$^3$,[21,24] resulting in a high local density of optical states (LDOS) in the gap (Fig. 1b and Fig. S1). In lateral directions $(x, y)$, the fields are strongly confined underneath the bottom facet of the nanoparticle of radius $R$ to lateral intensity full-width $\Delta_c \sim \sqrt{2Rd}/n$ with gap refractive index $n$[25] (Fig. 1c, Fig.S2). These cavity fields have a strong radiative component delivering high coupling efficiency to the far-field, $\eta \geq 0.5$.[21,24] A two-level emitter positioned in the gap experiences high LDOS and its emission is strongly enhanced ($\propto 1/V_c$). An emitter can thus be used to map these confined fields. However, it is challenging to precisely position a single-emitter within gaps of <5 nm with nm lateral resolution.

Various techniques[26] exist to control the location of a single emitter beyond random deposition: (i) chemical modification/functionalization of the substrate,[27] (ii) placement by scanning probes in scanning tunneling- or atomic force microscopes (AFM), (iii) electrostatic trapping,[28] (iv) capillary forces in solvent evaporation[29,30] and (v) host-guest chemistry.[31,32] Each of these methods suffer from inherent issues of randomness, coupled with often low-yield and difficult scalability. We achieve here deterministic bottom-up nanoassembly combining both organic and inorganic components, using deoxyribonucleic-acid origami (DNAo) nanotechnology.[33–38] A long single strand of DNA termed the scaffold is folded by the complementarity of base-pairs along hundreds of much shorter DNA 'staple strands'. These are designed to uniquely bind two or more sections of the scaffold together, while pinning different subcomponents to the staple strands yielding DNA 'breadboards' that can carry different functional elements.[39–47] We combine the two robust methods to form NPoM cavities with DNAo breadboard spacers. By precisely positioning a single-dye (Cy5) molecule at the center of the gap, we show the coherent coupling of cavity and emitter results in modulation of the cavity scattering spectrum. In addition, we map the LDOS with <3 nm precision by displacing the single-Cy5 molecule through the cavity in the lateral direction.

The DNAo is designed as a 2-layer plate (Fig. S3), each layer consisting of 24 helices having 128 to 149 base pairs[48] (see methods for detailed assembly procedure and S4, S5 and S16). The bottom layer has 4 thiol modifications on specific staple strands which are used to bind the origami onto the flat Au mirror. The top layer contains 6 poly-A (10 adenine bases on the 3') overhangs that can bind to the nanoparticle. The overhangs are designed to form a hexagon with the mid-point labelled (0,0) so that ssDNA-coated-nanoparticles hybridize to locate the center of the nanoparticle bottom facet there. The AFM images of these DNAo on a Au surface confirm a uniform size distribution and high yield assembly (Fig. 1e and Fig. S3). The ~2.5 nm diameter of each helix[49] sets the position of the overhangs from this center point (in nm) at $(x,y)$ of (0,5), (0,-5), (5,2.5), (5,-2.5), (-5,2.5) and (-5,-2.5). The zoomed AFM images (Fig. 1e) of individual structures show the clear features of these overhangs at the center and give the average thickness of the 2-layer plates as 4.5 ±0.3 nm (Fig.1f). The top plate is designed to bind Cy5-modified staples (chemical structure shown in Fig. 1d) which are 3' modified to locate them at coordinates divisible by 5 nm in the $y$-direction or internally modified stands to locate single Cy5 at ±2.5 and ±7.5 nm positions. Finally, 80 nm diameter Au nanoparticles functionalized with 5' thiol-modified 20x poly-T strands hybridize with the DNAo. The resultant assembly yields nanoparticles on a flat metal surface with the ultra-narrow gap (NPoM cavity) filled with DNA origami and a single-Cy5 molecule at the center (Fig. 1d). The optical emission of Cy5 from such a cavity is enormously enhanced due to the high LDOS within the gap.[50]

The robustness of assembled NPoM cavities are characterized for >350 nanoparticles using white-light dark-field nano-spectroscopy[51] (Fig. S6, dark-field image and spectrum). To first quantify the optical gap between the nanoparticle and Au mirror ($d$) and the refractive index ($n$), empty NPoMs without the Cy5 are constructed (Fig. 2a). Spectra of single NPoMs (Fig. 2b) show near identical peak positions, intensities and peak widths, further verifying the consistency of our robust nano-assembly.

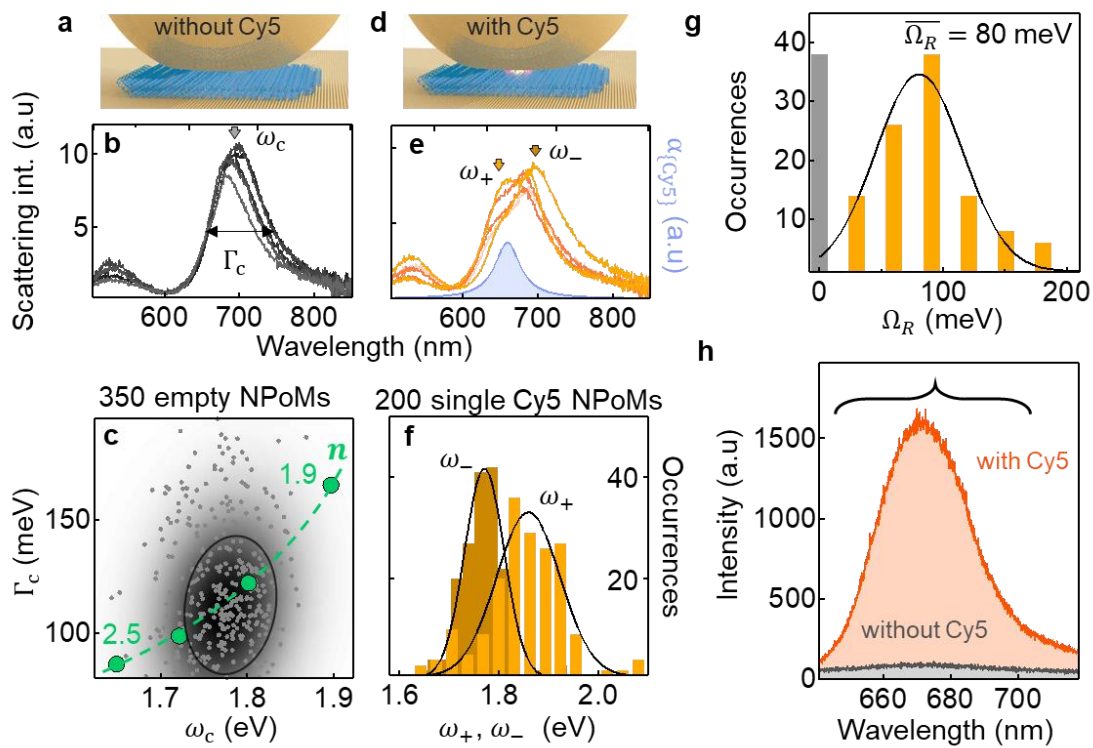

**Figure 2.** Characterization and coherent coupling with single-Cy5 in NPoM. (a,d) NPoM without and with Cy5 molecule in DNAo. (b,e) Experimental dark-field scattering of five individual NPoMs, with resonance peaks $\omega_{c,+,-}$ and linewidths $\Gamma_c$ marked. Absorption spectrum also shown (blue). (c) Cavity resonances *vs* linewidths for >200 NPoMs without (with) single Cy5 molecules. Background color map is kernel density matrix indicating distribution of NPoMs. Green dots are simulations for $n$= 1.9, 2.1, 2.3 and 2.5. (f) Frequency distribution of upper and lower split resonance energies with single Cy5 in each NPoM. (g) Distribution of calculated Rabi couplings extracted from (f), fraction of uncoupled NPoMs with $g < \Gamma_c$ highlighted in grey bar. (h) Optical emission from NPoMs with and without Cy5 molecule in the DNAo, laser at 633 nm. Curly bracket denotes spectral window used to integrate signal counts for enhancement factor below.

A characteristic infrared resonance peak ($\omega_c$, empty cavity) is identified at 1766±40 meV (702±18 nm) in the wavelength-dependent scattering spectra which corresponds to the NPoM coupled plasmon resonance. Spectral variations in the small peak around 530 nm indicate an average deviation in nanoparticle size of ±5 nm.[52] The AFM-measured thickness is used in electromagnetic simulations (Fig. S7) allowing extraction of the effective refractive index of the DNAo, which is modelled as an infinitely-wide sheet to simplify the geometry. The simulated coupled mode resonance positions and linewidths for different refractive indices in the gap ($n$=1.9, 2.1, 2.3 and 2.5) are plotted (green dots) along with the experimental data for >350 individual NPoMs (Fig. 2c). The statistical variation in $\omega_c$ and resonance full width at half maximum ($\Gamma_c$) fits $n$=2.15, in good agreement with previous studies of DNAo in closer proximity to Au.[53] Increases in nanoparticle size would shift $\omega_c$ to lower energy and increase $\Gamma_c$. Occasional linewidths <80 meV likely arise from the precise nano-geometry - the contact angle at the nanoparticle facet edge modulates the coupling strength between cavity and radiating antenna modes (see[54–56] for detailed analysis). The lack of correlation between ($\omega_c, \Gamma_c$) suggests that nanoparticle size is uncorrelated to the facet morphology. The robustness of $\omega_c$ and $\Gamma_c$ for such large samples of DNAo nanocavities verify that this DNAo-method allows for great control in the nanocavity formation, with uncertainties arising only from the fluctuations in nanoparticle shape and geometry.

Our NPoM cavity is designed with $2R$~80 nm so that the absorption and emission from the Cy5 molecule spectrally overlaps with $\omega_c$. Our simulations predict the field enhancements give large Purcell factors ($P_f \propto Q/V_c$), up to 4000 for single-Cy5 molecules embedded in the gap region[50] (Fig. 1c). The largest enhancements occur for Cy5 located at the center of the nanoparticle when the transition dipole is oriented vertically. Due to the unique plasmon mode hybridization in the NPoM nanocavity, this large field excitation is not quenched into non-radiative channels, as typically occurs when an emitter is placed close to a metal surface. Instead, in the NPoM nanocavity quenching is suppressed, leading to enhanced emission for the molecule that can be measured in the far-field [50].

By incorporating the single-Cy5 molecule into DNAo that assembles the NPoM (Fig. 2d), the optical scattering from the system is perturbed due to presence of single-Cy5 molecules. The resulting cavity resonance coupled with the single-Cy5 shows now two peaks which can be clearly resolved ($\omega_\pm$), for which we obtain the distributions from >200 NPoMs (Fig. 2f). Coherent coupling of the single-Cy5 absorption and emission (at $\omega_0$) with the detuned NPoM cavity does not quite reach the strong coupling regime of clear peak splittings (see discussion in Suppl.Info. S8 and S10), but still allows the coherent coupling strength $g$ to be extracted. This depends on the detuning $\delta = \omega_0 - \omega_c$ ~100 meV (Suppl.Info. S9). The emitter dephasing rate $\Gamma_0$ at room temperature is estimated to be 25 meV (~$k_B T$). The distribution of extracted coupling strengths for all NPoMs (Fig. 2g) gives a mean Rabi splitting $\overline{\Omega_R}$=80 meV, so that indeed $\Omega_R \sim (\Gamma_c + \Gamma_0)/2$ here. This is at the transition between weak and strong coupling regimes, while possessing low cooperativity values, $g^2/(2\Gamma_c\Gamma_0) <1$. The system approaches strong coupling, but the spontaneous emission rate still follows a Purcell-like dependence proportional to $g^2$,[57] valid for $g/\omega_0$ <0.1 (here $g/\omega_0$=0.03). Compared to our recent results in[21] the wider gap (4.5 nm *vs* 0.9 nm), but larger dipole strength of Cy5 (**μ**=10.1 D)[58] and twofold increase in damping (from the larger NP), gives a coupling rate $g$ only slightly smaller than in narrow gaps. Recent related efforts to position single molecules between plasmonic dimers with large gaps >10 nm suppressed the $g$ to small values and weakly enhanced emission rates.[43,59] The range of extracted Rabi splittings seen in Fig. 2g can originate from fluctuations in orientation and position of the Cy5, but also arises from a sub-population of bleached single Cy5 molecules, as we now discuss.

The presence of a single-Cy5 not only perturbs the cavity scattering but also enhances the optical emission from the Cy5 molecule. To measure the emission from individual NPoMs, we pump the cavity at 633 nm and collect all Stokes-shifted photons. In the absence of Cy5 molecules in the nanogap, the emission spectrum of a single NPoM is dominated by the surface enhanced Raman scattering (SERS) of DNAo and the inelastic light scattering (ILS) of electrons in Au from the plasmon resonance[60] (Fig. 2h, grey). The combined intensity of both phenomena vary across different NPoMs (Fig. S11). At relatively high pump powers >200 µW/µm², strong SERS signatures are identified at 1500 cm$^{-1}$ corresponding to cytosine, and at 1000 cm$^{-1}$ and 730 cm$^{-1}$ corresponding to adenine of DNA-origami. When a single-Cy5 is present at the center of the gap the emission from the NPoM is strongly enhanced (Fig. 2h, orange). Different NPoM constructs show only minor variations in peak emission wavelength and widths (Fig. S12a). The emission of Cy5 coupled to the plasmon mode enhances the decay rates and modifies the energy levels giving larger surface-enhanced fluorescence linewidths in comparison with the ensemble emission of molecules in solution (Fig. 2h, yellow).

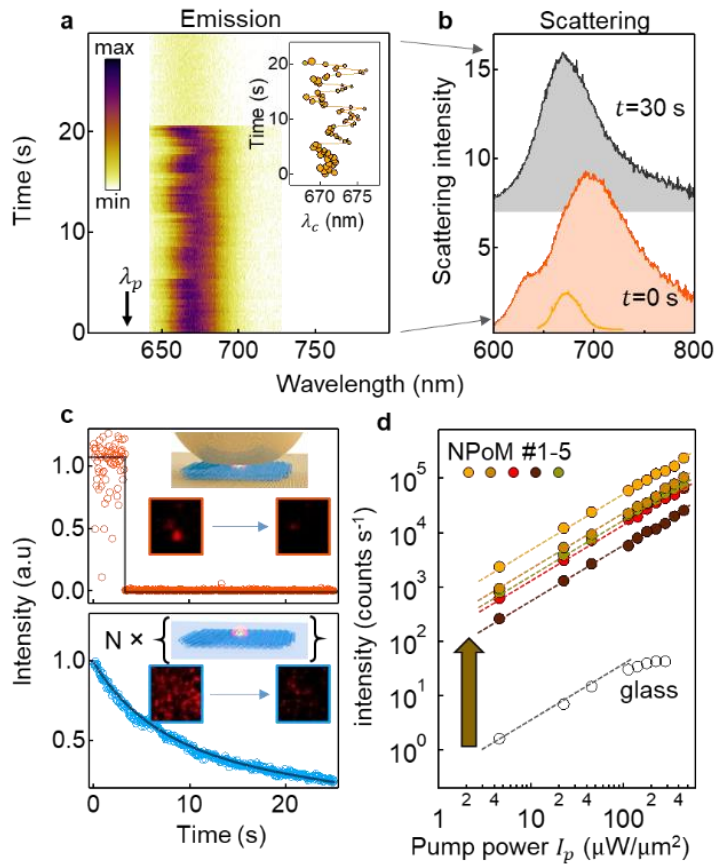

**Figure 3.** Emission from NPoM with single-Cy5. (a) Time-series emission spectra from a NPoM with single-Cy5 at the gap center. Inset: extracted variation in peak $\lambda$ and intensity (dot size) as a function of time. (b) Scattering spectra from the NPoM before ($t$=0 s) and after bleaching ($t$=30 s) the single-Cy5, with initial emission spectrum (orange). (c) Intensity over time extracted from fluorescence microscopy images (shown in inset) for single-Cy5 in NPoM (top) and ensemble Cy5 embedded in DNAo on glass (bottom). (d) Emission intensity as a function of pump power in five individual NPoMs, and for comparison a typical DNAo on glass.

Time series scans of the emission from NPoMs containing a single Cy5 show variations in peak position and intensity on timescales of seconds (Fig. 3a and Fig. S12b). The emission suddenly bleaches after a certain time, leaving only weak ILS and SERS from the NPoM ($t$>21 s in Fig. 3a) which is the same for the DNAo without Cy5 (Fig. 2h, grey). This confirms the presence of only one Cy5 molecule in each NPoM, as previously demonstrated for single emitters within DNAo inside plasmonic dimers.[43,45] Scattering spectra obtained from the NPoM before (Fig. 3b, bottom) and after (Fig. 3b, top) the bleaching of the single-Cy5 show the expected collapse in the splitting. Intensity traces from fluorescence microscopy images of NPoMs also show these step-like features, in complete contrast to ensemble Cy5s on glass which show the gradual irreversible bleaching of molecules. We note that chemically binding Cy5 to DNA stands has been shown to increase the photo-stability of the molecules.[61] The average bleaching time of Cy5-DNAo on glass is $\bar{\tau}_b$=8.5 s, whereas in these NPoMs we find $\bar{\tau}_b$ > 13 s, giving an additional two-fold increase in photo-stability. With NPoMs providing local intensity enhancements of $I_{\text{gap}}$~ 2500 in the gap (Fig. 1c) and $\tau_b \propto 1/I_{\text{gap}}$, the photobleaching is actually suppressed by more than three orders of magnitude.[62] The total number of photons emitted from the single dye in each NPoM before it bleaches is estimated to be 4×10⁶, resulting in enhancement of total photon counts by factors >1000 times compared to each Cy5 in DNAo on glass. This enhancement results from combined effects of enhanced radiative emission rates, better light collection from the NPoM antenna, and suppressed bleaching rates.[46,63] These behaviors fully corroborate our evidence for single molecule emission.

To estimate the emission enhancement of Cy5 molecules in NPoMs we performed emission experiments for different pump powers (Fig. S13). Emission from an ensemble of Cy5 molecules on glass exhibits saturation and bleaching at >100 µW/µm² (Fig. 3d, black curve). The emission is now integrated over a range of wavelengths (curly bracket Fig. 2h) and normalized to the counts from Cy5s in DNAo on glass considering the excitation and collection efficiencies (Suppl.Info. S14), to give the enhancement factor (EF). We find the radiative emission rate ($\gamma_e/\gamma_0$) is enhanced by EF > 2×10³. The coupling strength estimated from this single-Cy5 enhancement[10] using $g^2 = P_f \gamma_0 \Gamma_c (1 + 2\delta/\Gamma_c)/4$ (with radiative linewidth $\hbar/\gamma_0$=1 ns[64]) gives values $g$ ~50 meV which agree with those from the cavity linewidth in scattering (Fig. 2g). The emission intensity from a Cy5 coupled to a single NPoM shows linear scaling with excitation power density in the range 4-400 µW/µm² (Fig. 3d). We estimate photon populations in the NPoM < 0.1, well below the critical cavity photon population[65] for non-linear effects $\Gamma_c^2/(2g^2)$~2. Pumping at higher excitation power densities instead gives irreversible photobleaching of the Cy5, before saturation of the excited state population can be reached. All subsequent measurements are thus conducted at excitation power densities of 50 µW/µm². At these powers we do not see the collapse of the DNA origami core as seen in prior work at >500 µW/µm².[45,66,67]

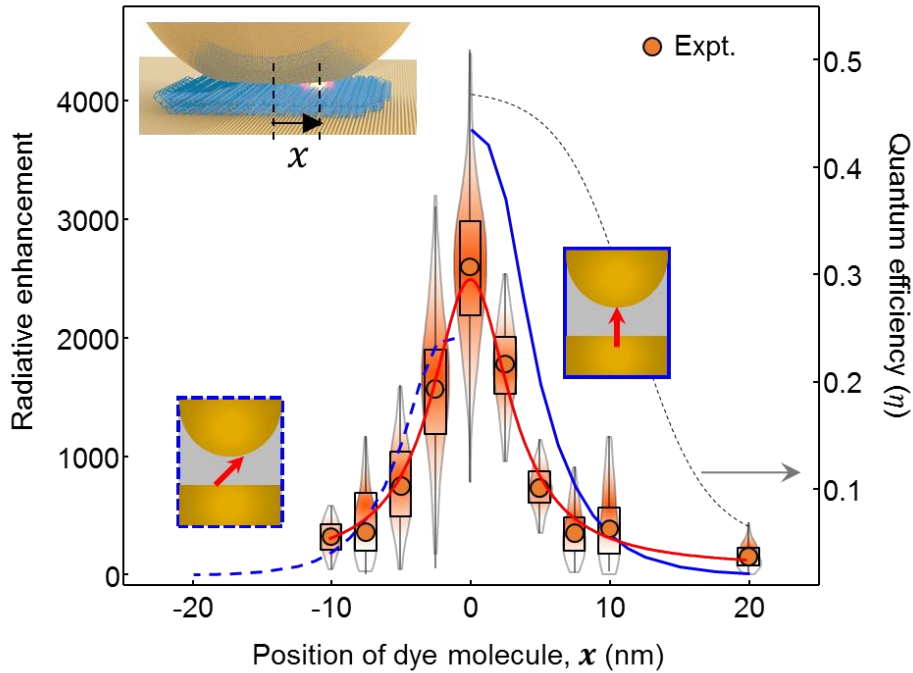

**Figure 4.** Mapping the LDOS. Experimental variation in enhancement of emission intensity from NPoMs when laterally displacing the position of single dye molecule (orange dots). Normalized to the emission of a single Cy5 in DNAo on glass, grey box indicates standard error, statistical variations in emission intensity shown as violin plots. Solid red line is Gaussian fit to the experimental data. Theoretically calculated emission enhancements are shown for vertical ($90^0$) dipole (solid blue) and slanted dipole ($45^0$) (dashed blue). Calculated quantum efficiency for vertical dipole is shown as dashed grey line.

The EF is measured for >100 individual NPoMs for different assembled origamis. In successive designs, the spatial position $x$ of the single-Cy5 is systematically scanned laterally within the gap (Fig. 4). As the Cy5 is moved towards the center of the gap ($x$=0) the emission intensity increases monotonically, evidencing that the centre of the NP on the DNAo is correctly defined within $\pm 1$ nm, and that the optical field in the gap has a spatial fullwidth $\Delta_{\text{expt}}$ = 6.5±2 nm. This measured intensity profile is similar to that from simulations $\Delta_c \sim 8$ nm for facets $w$<10 nm (blue lines Fig. 4, Fig. S2) as well as the analytical estimate $\sqrt{2Rd}/n \sim 9$ nm. The statistical variation of EFs for each design (shown with violin plots around each point) show the deterministic assembly achieved here. To the best of our knowledge, this is the first time the optical field of plasmonic nanogaps is mapped deterministically with single-molecules. We note that this measurement also confirms the accuracy of the single-molecule Cy5 positioning within DNAo to <1.5 nm (as this would otherwise blur out the spatial fullwidth further, see Supply.Info. S15), which is surprising. Compared to near-field scanning microscopies with metal tips that strongly perturb the confined optical mode, the present technique for measuring cavity optical fields using deterministic placement through DNA origami is minimally invasive.[68,69]

The experimental data matches full 3D electromagnetic simulations for the radiative enhancement with two different dipole orientations ($90^0$ and $45^0$ shown solid and dashed). We find a best fit for dipole orientations of $65^0 \pm 15^0$. Different DNA-origami folding results in slightly different dipole orientations, and partial melting of the double-stranded DNA together with slight imprecision in nanoparticle placement yields the uncertainty in emitter position.

It is evident from this data that an emitter in a plasmonic nanocavity does not quench when placed in the vicinity (<5 nm) of the two Au surfaces. This result is a consequence of the enhanced emission rate outstripping the absorption rate as we showed earlier.[50] Instead its emission rate is strongly enhanced when moved towards the center of the nanocavity. It is important to note that our geometry is completely different to the quenching observed when emitters are placed close to a *single* Au surface.

Integrating robust NPoM constructs with high precision DNA origami techniques, we show that a single-molecule deterministically positioned at the center of a nanocavity interacts coherently with confined optical fields ($\Omega_R \sim 80$ meV) producing splitting near the weak-to-strong coupling regime (with cooperativity~1). Modification of the scattering spectra before and after bleaching each single-Cy5 shows a type of energy switching at room temperature that requires only zJ to break a bond, which although currently irreversible can be now explored in photochromic and electro-optic molecules. The optical emission from each single-molecule can be enhanced by >$10^3$ showing that our NPoM constructs allow for strong fluorescence despite close proximity of the dye to metal surfaces. Further, by systematically moving the position of molecule through the cavity, we map the local field confinement with high accuracy. We believe that such robust systems are ideal for studying room-temperature single-molecule nano-optics, and have the potential for a variety of technological implementations.

**METHODS: DNA Origami and NPoM assembly.** The origami are folded in a 14 mM $MgCl_2$, 1x TE Buffer using a 7560-base long single stranded viral DNA scaffold isolated from M13mp18 derivative (*Tilibit nanosystems*) at a concentration of 10 nM and a staple concentration of 100 nM (i.e. 10:1 staple: DNA). The folding is carried out using an annealing cycle that slowly cools the solution from 65 °C to 36 °C over a period of 23 hours followed by holding at 4 °C. Once the cycle is complete, the solution is filtered through a 100 kDa *Amicon* filter in a 2 mM $MgCl_2$, 0.5x TBE washing the buffer thrice.

The origamis are then allowed to functionalize on a gold mirror substrate overnight in a 11 mM $MgCl_2$, 0.5x TBE buffer. Finally, 2$R$=80 nm gold particles (*BBI solutions*) functionalized with 5' thiol modified 20x poly T strands are allowed to hybridize with the origami for at least 30 minutes, prior to being rinsed with milliQ water and blown dry with nitrogen.

The dye molecule positions are set by their binding to the overhangs. In principle it is possible to have finer control over the lateral position by using rotation along the helix. In our case, the helices containing the dyes are designed to be approximately at the same rotation around the helix, although the ±2.5/7.5 nm positions are reversed – the staples are approximately pointing in the opposite vertical direction to those at {0,5,10nm}. It should thus be noted that the vertical positions are potentially different, around (0,0,0) and (0,2.5,-2.5).

**Simulations.** The 3D numerical simulations are performed using Lumerical FDTD Solutions v8.12. The Au NP was modeled as a sphere or a truncated sphere (to model different facet sizes) of different $R$ (70-90 nm) on top of an infinite dielectric sheet of variable $n$=1.9-2.5 and thickness of 4.5 nm. Underneath this sheet, a thick gold layer is placed to replicate the experimental nanoparticle-on-mirror geometry. The dielectric function of gold is taken from Johnson and Christy. The nanoparticle was illuminated with a p-polarized plane wave (TFSF source) from an angle of incidence of $\theta_i$=55°. The inbuilt sweep parameter is used to include the incident wavelengths ranging from 500 to 900 nm. The scattered light at each wavelength was then collected within a cone of half angle $\theta_c$ = 55° based on the numerical aperture of the objective.

ASSOCIATED CONTENT

**Supporting Information**.

AUTHOR INFORMATION

**Corresponding Author**

* Prof Jeremy J Baumberg, jjb12@cam.ac.uk

**Notes**

The authors declare no competing financial interest.

ACKNOWLEDGMENT

We acknowledge financial support from EPSRC Grants EP/G060649/1, EP/K028510/1, EP/L027151/1, ERC Grant LINASS 320503 and ERC Consolidator Grant (DesignerPores 67144). R.C. acknowledges support from the Dr. Manmohan Singh scholarship from St. John's College. F.B. acknowledges support from the Winton Programme for the Physics of Sustainability.